\documentclass[journal,12pt,onecolumn,draftclsnofoot,]{IEEEtran}

\usepackage{amsmath}
\usepackage{amsfonts}
\usepackage{amssymb}
\usepackage{multicol}
\usepackage{verbatim}
\usepackage{balance}
\usepackage{epstopdf}
\usepackage{epsfig,color,algorithm}
\usepackage{pdflscape}
\usepackage{rotating}
\usepackage{adjustbox}
\usepackage{booktabs, threeparttable}
\usepackage{array}
\usepackage{multirow}
\usepackage{graphicx}
\usepackage{lipsum}% example text
\usepackage[table,xcdraw]{xcolor}
\usepackage{algorithmic}
\newcolumntype{L}[1]{>{\raggedright\arraybackslash}p{#1}}

\ifCLASSOPTIONcompsoc
  \usepackage[nocompress]{cite}
\else
  \usepackage{cite}
\fi
\ifCLASSINFOpdf

\else

\fi
\begin{document}
\title{Energy-Efficient MAC for Cellular IoT: State-of-the-Art, Challenges, and Standardization}
\author{
\IEEEauthorblockN{Syed Waqas Haider Shah, \textit{Student Member, IEEE}, Adnan Noor Mian, \textit{Member, IEEE}, Adnan Aijaz, \textit{Senior Member, IEEE}, Junaid Qadir, \textit{Senior Member, IEEE}, and Jon Crowcroft, \textit{Fellow, IEEE}}
}
\maketitle
\long\def\symbolfootnote[#1]#2{\begingroup%
\def\thefootnote{\fnsymbol{footnote}}\footnote[#1]{#2}\endgroup}
\symbolfootnote[0]{\hrulefill \\
Syed Waqas Haider Shah and Jon Crowcroft are with the Computer Lab, University of Cambridge, 15 JJ Thomson Avenue, Cambridge, UK CB3 0FD (\{sw920, jon.crowcroft\}@cl.cam.ac.uk). Syed Waqas Haider Shah is also with the Electrical Engineering Department, Information Technology University, Lahore 54000, Pakistan (waqas.haider@itu.edu.pk).

Adnan Noor Mian is with the Computer Science Department, Information Technology University, Lahore 54000, Pakistan (adnan.noor@itu.edu.pk).

Adnan Aijaz is with the Telecommunications Research Laboratory, Toshiba Research Europe Ltd., Bristol, UK (adnan.aijaz@toshiba-trel.com).

Junaid Qadir is with the Electrical Engineering Department, Information Technology University, Lahore 54000, Pakistan (junaid.qadir@itu.edu.pk).
}

\maketitle

\begin{abstract}
In the modern world, the connectivity-as-we-go model is gaining popularity. Internet-of-Things (IoT) envisions a future in which human beings communicate with each other and with devices that have identities and virtual personalities, as well as sensing, processing, and networking capabilities, which will allow the developing of smart environments that operate with little or no human intervention. In such IoT environments, that will have battery-operated sensors and devices, energy efficiency becomes a fundamental concern. Thus, energy-efficient (EE) connectivity is gaining significant attention from the industrial and academic communities. This work aims to provide a comprehensive state-of-the-art survey on the energy efficiency of medium access control (MAC) protocols for cellular IoT. we provide a detailed discussion on the sources of energy dissipation at the MAC layer and then propose solutions. In addition to reviewing the proposed MAC designs, we also provide insights and suggestions that can guide practitioners and researchers in designing EE MAC protocols that extend the battery life of IoT devices.  Finally, we identify a range of challenging open problems that should be solved for providing EE MAC services for IoT devices, along with corresponding opportunities and future research ideas to address these challenges.

\end{abstract}

\IEEEpeerreviewmaketitle


\section{Introduction}
\begin{figure*}[]
\centering
{\includegraphics[width=6in]{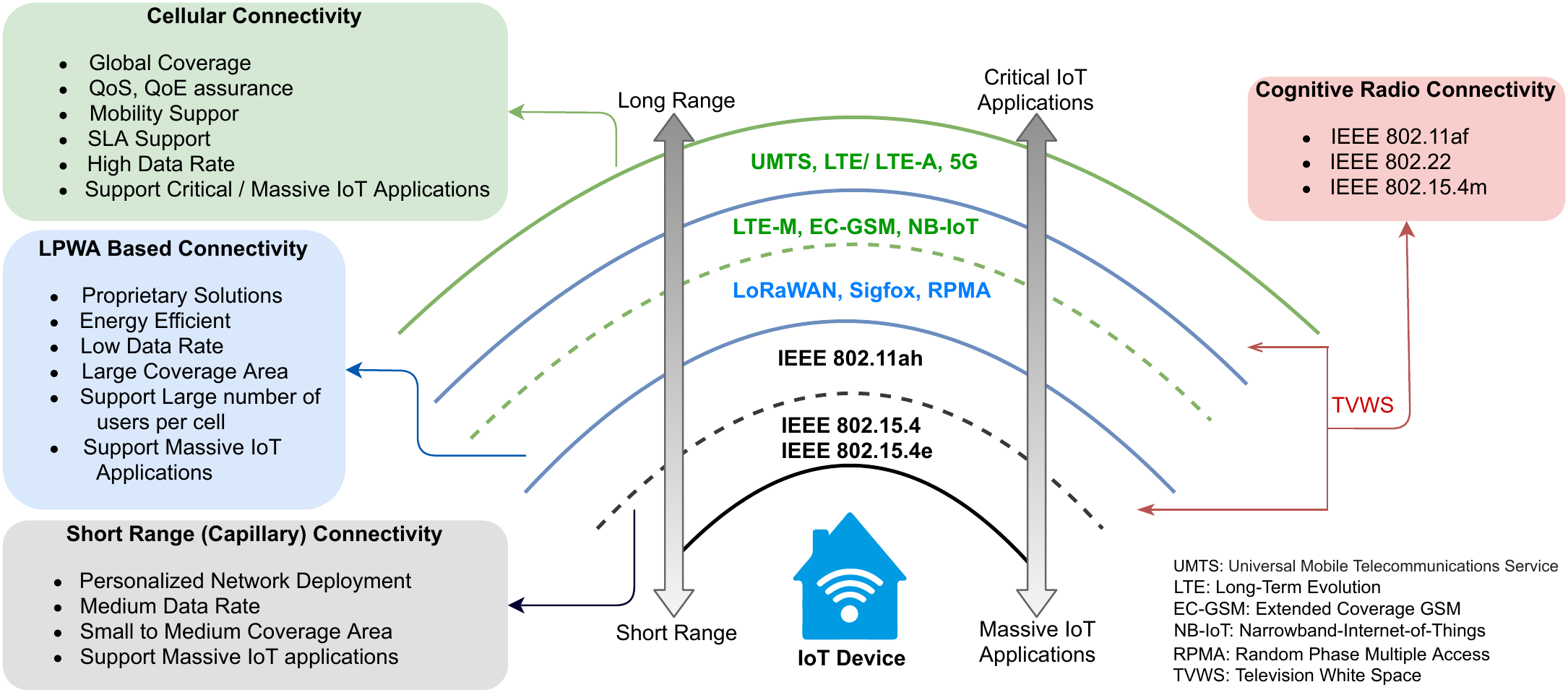}}
\caption{IoT wireless connectivity ecosystem: covers licensed and unlicensed spectrum operating networks.}
\label{ecosystem}
\end{figure*}

Internet-of-Things (IoT) enables a paradigm shift from a manual approach to an automated system. It manages resources in an efficient way that not only improves productivity but also cost and time efficiency. The Internet, which is the backbone of the IoT, has grown exponentially since the last decade. The ``connectivity-as-we-go" model becomes a reality with the help of wireless Internet connectivity. As the popularity of IoT grows, it attracts more sectors of society to become automated. IoT applications are generally categorized into massive IoT (m-IoT) and critical IoT (c-IoT). The m-IoT applications are generally delay-tolerant, such as smart home, personal IoT, and smart metering. Whereas, c-IoT applications are delay-sensitive, like smart healthcare, industrial automation, and asset tracking \cite{jammes2016internet}. Although IoT integration in multiple sectors brings many benefits, it also creates connectivity and energy challenges. There are various wireless networks available, which are broadly divided into licensed and unlicensed. Licensed networks, such as cellular networks, render numerous benefits over unlicensed networks, as shown in IoT wireless connectivity ecosystem in Fig. \ref{ecosystem}. The cellular network attributes, such as quality-of-service (QoS) and quality-of-experience (QoE) assurance, mobility and service-level-agreement (SLA) support, high data rates, and global coverage, make them a better candidate not only for m-IoT applications but also for c-IoT applications. However, an IoT device consumes more energy when it is connected to a cellular network. It is mainly due to the complex and power-hungry network infrastructure of cellular networks. Furthermore, IoT devices are mostly battery-operated. Thus, to make a device last longer, it is necessary to save as much power as possible in its operation. 

Energy-efficient (EE) IoT paradigm (also known as green IoT) aims at reducing the power consumption of an IoT device with the help of efficient and sustainable technologies. The maximum energy consumption of an IoT device associates with its physical and medium access control (MAC) layer operations \cite{galmes2018optimal}. MAC designs for Cellular networks were primarily designed for conventional cellular communication. These MAC protocols are unable to address large-scale concurrent channel access in an IoT network. Recently, the 3rd generation partnership project (3GPP), which develop protocols for mobile telecommunications, propose various modifications in cellular MAC in its Release 12-16. These modifications, which ensure the EE operation of IoT devices in cellular networks, help realize the green cellular IoT.  To this end, we survey various EE MAC protocols and cellular standards specifically designed for green IoT.

Due to its paramount importance, the energy efficiency of MAC protocols has been widely studied in the literature. The authors in \cite{rajandekar2015survey} survey the technical challenges associated with the MAC layer for machine-type communication (MTC). They discuss some of the key elements of the MAC layer designs, including throughput, latency, scalability, and energy efficiency. Although the EE aspect of the MAC layer protocols for IoT is discussed, the authors do not discuss cellular IoT. Another article surveys the MAC protocols for IoT \cite{oliveira2019mac}. The authors classify MAC protocols according to the IoT connectivity scenario. They discuss MAC protocols for IoT in both licensed and unlicensed networks. This survey also lacks a discussion on the EE aspects of the protocols. Similarly, another study surveys the suitability of the current random access procedure of cellular networks for IoT \cite{laya2014random}. The authors present the case for the unsuitability of the conventional cellular networks for the EE IoT connectivity. Various other surveys exist as well, which study the EE MAC protocols \cite{luitel2018energy,ghafoor2020mac,li2019survey,8536384}. However, their scope is limited to one network connectivity, and none of them discusses the EE MAC protocols for cellular IoT. For example, the authors in \cite{ghafoor2020mac} study MAC protocols for terahertz communication only. The authors in \cite{li2019survey} only discuss EE MAC protocols for wireless multimedia sensor networks. Another survey \cite{luitel2018energy} provides EE MAC protocols for cognitive radio sensor networks. Similarly, the authors in \cite{8536384} present a survey on the EE narrowband IoT (NB-IoT). Nevertheless, there is a gap in the literature regarding the energy efficiency aspects of MAC layer protocols for cellular IoT, as shown in Table \ref{literature_sur}.
\begin{table*}[]
\centering
\caption{Literature Review Summary (Yes = Included, No = Not Included, Partially = Partially Included)}
\label{literature_sur}
\resizebox{\textwidth}{!}{\begin{tabular}{|c|c|l|l|l|l|l|}
\hline
\multirow{2}{*}{\textbf{\begin{tabular}[c]{@{}c@{}}Survey\\ Paper\end{tabular}}} & \multirow{2}{*}{\textbf{Year}} & \multicolumn{1}{c|}{\multirow{2}{*}{\textbf{Network}}}                           & \multicolumn{1}{c|}{\multirow{2}{*}{\textbf{EE-MAC}}} & \multicolumn{2}{c|}{\textbf{\begin{tabular}[c]{@{}c@{}}Energy\\ Consumption\end{tabular}}}                      & \multicolumn{1}{c|}{\multirow{2}{*}{\textbf{Brief Summary}}}    \\ \cline{5-6} &                                & \multicolumn{1}{c|}{}       & \multicolumn{1}{c|}{}                                 & \multicolumn{1}{c|}{Sources} & \multicolumn{1}{c|}{\begin{tabular}[c]{@{}c@{}}Saving\\ Techniques\end{tabular}} & \multicolumn{1}{c|}{} \\ \hline
\textit{\begin{tabular}[c]{@{}c@{}}Rajandekar\\ et al.\cite{rajandekar2015survey}\end{tabular}}    & 2015                           & Non-cellular                                                                     & No                                                    & No                           & No                                                                               & \begin{tabular}[c]{@{}l@{}}The authors perform a survey of the technical challenges and the\\ requirements of the MAC layer for M2M communication. The s-\\urvey outlines key requirements of MAC layer design for M2M,\\ such as throughput, latency, scalability, energy efficiency, and co-\\existence, followed by a taxonomy of M2M MAC protocols.\end{tabular}                                                                                                                                                                  \\ \hline
\textit{\begin{tabular}[c]{@{}c@{}}Luitel\\ et al. \cite{luitel2018energy}\end{tabular}}                 & 2018                           & \begin{tabular}[c]{@{}l@{}}Cognitive\\radio sensor\\network\end{tabular}         & Partially                                             & No                           & No                                                                               & \begin{tabular}[c]{@{}l@{}}The survey provides a qualitative comparison of the EE MAC pr-\\otocols for CRSN followed by a discussion on the design issues\\ for the MAC layer protocols. Later, the authors also present sev-\\eral research challenges related to MAC designs for CRSN.\end{tabular}                                                                                                                                                                                                                                    \\ \hline
\textit{\begin{tabular}[c]{@{}c@{}}Popli\\ et al. \cite{8536384}\end{tabular}}                  & 2018                           & NB-IoT                                                                           & Partially                                             & No                           & No                                                                               & \begin{tabular}[c]{@{}l@{}} The survey highlights various design challenges, which are acting\\
as limiting factors for the success of the NB-IoT. It also t presents\\two EE techniques named EE adaptive health monitoring system\\ and zonal thermal pattern analysis, for establishing green IoT appl-\\ications.\end{tabular}                                                                                                                                                         \\ \hline
\textit{\begin{tabular}[c]{@{}c@{}}Oliveira\\ et al. \cite{oliveira2019mac}\end{tabular}}               & 2019                           & \begin{tabular}[c]{@{}l@{}}Cellular +\\ Non-cellular\end{tabular}                & No                                                    & No                           & No                                                                               & \begin{tabular}[c]{@{}l@{}}This survey discusses MAC protocols for both licensed and unlice-\\nsed wireless networks, including LPWA technologies and cellular\\standards for IoT. It also provides a comparative study for each gr-\\oup of protocols to provide insights and a reference study for IoT\\applications, considering their characteristics and limitations.\end{tabular}                                                                                                                                                     \\ \hline
\textit{\begin{tabular}[c]{@{}c@{}}Shu Li\\ et al. \cite{li2019survey}\end{tabular}}                 & 2019                           & \begin{tabular}[c]{@{}l@{}}Wireless\\multi-media\\sensor\\network\end{tabular} & Partially                                             & Partially                    & No                                                                               & \begin{tabular}[c]{@{}l@{}} The survey provides unique features and the requirements for desi-\\gning the WMSN, followed by a summary of the possible solutions.\\It also studies several MAC protocols for WMSN in terms of their\\ prioritization and service differentiation mechanisms. It also provid-\\es EE and QoS-enabled routing protocols for the WMSN.\end{tabular}                                                                                                       \\ \hline
\textit{\begin{tabular}[c]{@{}c@{}}Ghafoor\\ et al. \cite{ghafoor2020mac}\end{tabular}}                & 2020                           & Terahertz                                                                        & No                                                    & No                           & No                                                                               & \begin{tabular}[c]{@{}l@{}}The survey investigates the key features of the Terahertz band that\\needs to be regarded for designing efficient MAC protocols. It also\\ provides design challenges and their possible remedies for MAC pr-\\otocols in applications at macro and nano scales. It also compares\\MAC protocols based on channel access, network topologies, and\\link establishment procedures followed by a discussion on the open\\research challenges for Terahertz MAC protocols.\end{tabular}                                         \\ \hline
\textit{\begin{tabular}[c]{@{}c@{}}D. Jie\\ et al. \cite{ding2020iot}\end{tabular}}                  & 2020                           & \begin{tabular}[c]{@{}l@{}}Cellular + \\ Non-cellular\end{tabular}               & No                                                    & Partially                    & No                                                                               & \begin{tabular}[c]{@{}l@{}}The survey highlights fundamental challenges and bottlenecks of s-\\everal emerging connectivity technologies, like compressive sensing,\\ nonorthogonal multiple access, and machine learning-based random\\access. It also provides a detailed discussion on the possible solut-\\ions to address these challenges. Lastly, it classifies IoT applications\\for different technical domains and a discussion on the suitable IoT\\ connectivity technologies.\end{tabular}                                           \\ \hline
\textit{This Paper}                                                              & 2021                           & Cellular                                                                         & Yes                                                   & Yes                          & Yes                                                                              & \begin{tabular}[c]{@{}l@{}}This survey identifies various sources of energy dissipation at the\\MAC layer and their possible solutions. It also raises several ques-\\tions on the suitability of various MAC protocols for EE IoT conn-\\ectivity in future cellular networks. Further, it provides a quantitative\\ comparison between proprietary LPWA technologies and cellular\\standards for IoT in the context of their energy consumption. Later, it\\ identifies various challenges for possible future research directions.\end{tabular} \\ \hline
\end{tabular}}
\end{table*}

The main objective of this paper is to fill this gap by providing a comprehensive survey on the EE MAC designs for green cellular IoT. We identify various sources of energy dissipation at the MAC layer and their possible solutions. We raise several questions on the suitability of various MAC protocols for EE IoT connectivity in future cellular networks. Moreover, a quantitative comparison between different proprietary low power wide area (LPWA) technologies and cellular standards for IoT is provided in the context of their energy consumption. Lastly, we identify various challenges for possible future research direction.

The rest of the paper is organized as follows. Section II explains the energy consumption at the MAC layer; Section II-A provides sources of energy dissipation, and Section II-B presents techniques for minimizing energy consumption. Section III presents the EE MAC protocols for cellular IoT. Section IV provides the EE cellular standards especially designed for IoT connectivity. Section V presents cellular community efforts in providing EE access to massively deployed IoT devices. Section VI introduces open challenges for future research, and Section VII concludes the paper.

\section{Energy Consumption at the MAC Layer}
\begin{figure}[ht]
\centering
{\includegraphics[width=3.1in]{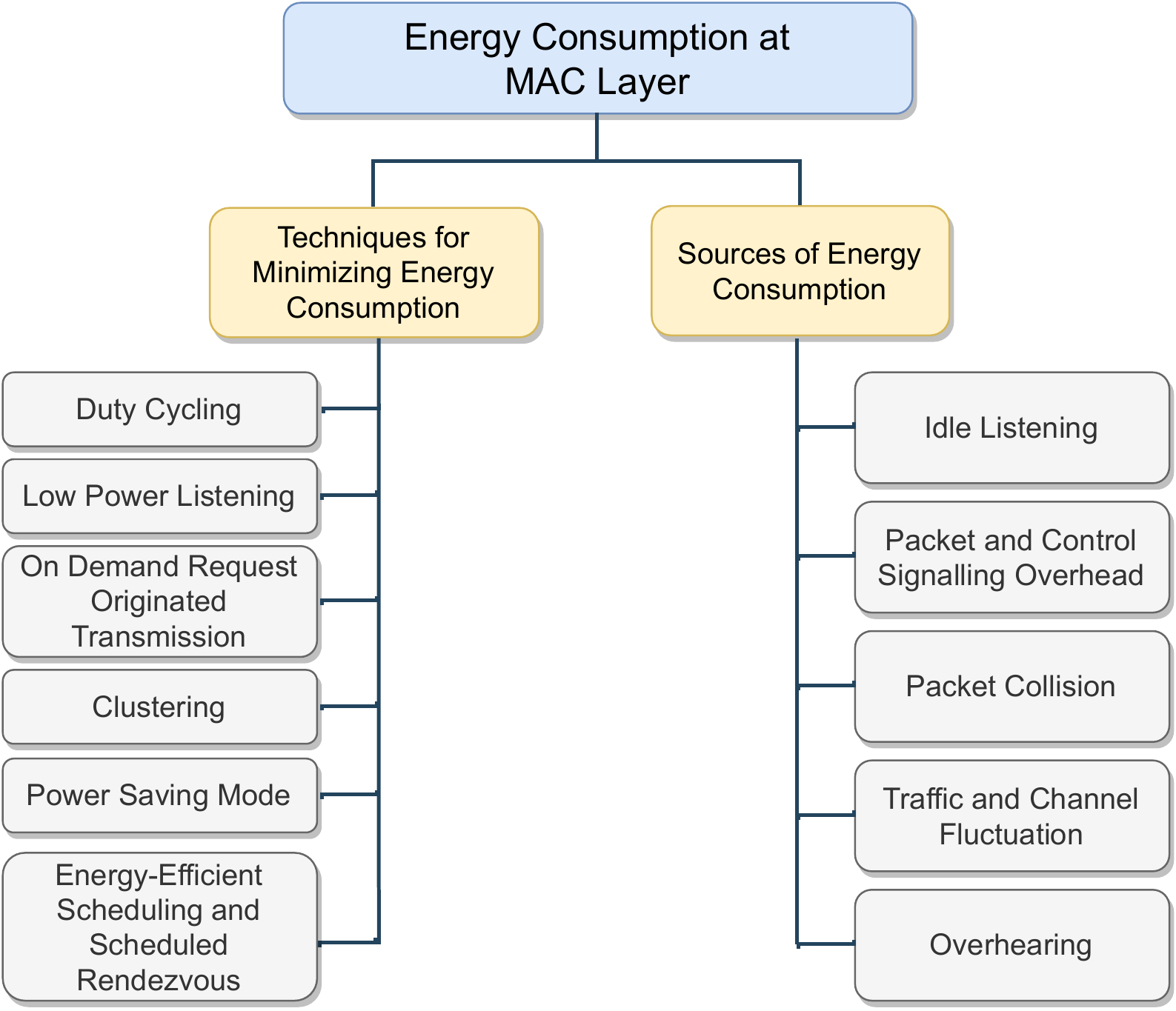}}
\caption{Taxonomy of energy consumption sources at the MAC layer and approaches to minimize the energy consumption. }
\label{MAC_layer}
\end{figure}

In this section, we identify sources of energy dissipation and the possible techniques for minimizing energy consumption at the MAC layer for IoT connectivity, as shown in Fig. \ref{MAC_layer}.
\subsection{Sources of Energy Dissipation at the MAC Layer}
The majority of IoT devices are battery-powered, and the energy consumption occurs in two basic operational modes; active mode and sleep mode. In an active mode, the radio of a device remains on, and the power consumption of the device depends on the coverage scenario of the network in which the device is connected. If the network coverage is good, a device requires less power to establish a connection and data transfer. However, in a poor network coverage scenario, the power consumption of a device is high due to the re-transmission of the lost packets and because of the low order modulation schemes for a better signal-to-noise ratio. In a sleep mode, the power consumption of a device is comparatively low because the radio of the device remains off for a longer period of time. Now, we will discuss some common sources of energy dissipation at the MAC layer, as shown in Fig. \ref{MAC_layer}.

\subsubsection{Idle Listening}
Idle listening leads to the undesirable energy consumption of an IoT device. It happens when a device expects to receive a packet and keeps its radio on, but the packet was not sent from the transmitting device. Similarly, when a device transmits a packet to a device that is in sleep mode, the transmitting device keeps its radio on for an acknowledgment. It can happen due to unsynchronized sleep-wake scheduling between transmitting and receiving devices. It can be avoided using the power-saving mode (PSM), which will be discussed later in the section.
\subsubsection{Packet and Control Signaling Overhead}
In cellular networks, devices and the BS share preliminary information to keep the network updated; this is called control information. Different protocols use control signaling to avoid collisions, such as ready to send /clear to send packets. These packets do not contain any data but consume power, which is undesirable in some cases. Moreover, the data payload in MTC communication is small, and the control signaling is overwhelmingly large.
\subsubsection{Packet Collision}
Packet collision occurs when multiple devices transmit a packet on a shared channel simultaneously. When a collision occurs, the transmitted packet is lost, and devices need to retransmit the packet. This way, devices consume more energy, first in collision avoidance and then in retransmitting the data. In cellular IoT networks where connected devices are in large number, packet collision becomes more critical. 
\subsubsection{Traffic and Channel Fluctuation}
In cellular networks, channel conditions change rapidly, and channel estimation is required to propose an efficient system model. The channel state information (CSI) provides the overall behavior of the channel on a time/frequency scale, and it needs to be updated periodically. Resource scheduling for the devices in a cellular network is also based on channel fluctuation \cite{zhu2011instantaneous}. When the channel or traffic conditions change rapidly, devices need to keep their radio on for expected packets arrival; thus, they consume an extra amount of energy.
\subsubsection{Overhearing}
Overhearing is a phenomenon in which a device receives a packet that was not sent for it. Due to this, devices end up consuming extra energy. One of the main reasons for overhearing is continuous listening of the channel to find an idle space for transmission. It becomes more critical in cellular IoT networks due to a large number of connected devices. 

\subsection{Techniques for Minimizing Energy Consumption}
\subsubsection{Duty Cycling}
Generally, an IoT device generates a small amount of sensed information or control signaling information. In most applications, an IoT device generates a packet every few seconds and remains idle otherwise. Power consumption when an IoT device is in idle mode is of no use and should be reduced. To do so, the duty cycling mechanism can be used that allows an IoT device to follow a sleep-awake schedule. Energy consumption of an IoT device can be characterized in four operational modes; sleep, idle, transmit, and listen. In transmit, listen, and idle modes, maximum energy consumption is due to the radio being on. Whereas, when the radio is off (in sleep mode), a small amount of energy is consumed. Turning the radio off of an IoT device when there is no active communication can save up to 50\% energy \cite{ye2002energy}. However, there is a tradeoff between energy consumption and transmission latency in duty cycling. Although a small duty cycle can save energy, it will increase the transmission latency. Multiple MAC protocols for duty cycling are available in the literature \cite{du2007rmac,buettner2006x,sun2008ri}. A survey on duty cycle protocols for wireless sensor networks is also available in \cite{alfayez2015survey}. The authors in this survey characterize various duty-cycled MAC protocols in synchronous and asynchronous approaches.
\subsubsection{Low Power Listening (LPL)}
In LPL, an IoT device wakes up without any synchronization procedure for a short time to check the channel status \cite{panta2012efficient}. It uses a channel polling mechanism, in which if a device senses a busy channel, it remains in an active mode for data reception, and the rest of the devices in the network go to sleep mode. When a device is transmitting, it uses a long preamble with every packet to ensure that the receiver is awake when the packet arrives. The preamble size is adjustable according to the traffic situation in the network. LPL is a suitable technique for MTC communication in IoT networks, as it requires low implementation complexity and can afford small duty cycling. However, its performance can be compromised due to the false wake-up of a device. In noisy environments, channel-induced noise may be detected as a channel activity, which causes a device to remain in an active mode. The energy level detection is a critical factor in LPL performance, and the authors in \cite{sha2013energy} proposed an adaptive energy detection protocol. In this protocol, an IoT adaptively adjusts its energy threshold level to avoid false wake-up. Hwang et al. The authors in \cite{hwang2014taxonomy} provide a detailed survey on taxonomy and evaluation of LPL protocols for the suitability of IoT networks.
\subsubsection{On-Demand Request Originated Transmissions}
In an IoT network, devices are generally deployed on a large scale, including high-rise buildings and hard to reach areas. These devices, which are mostly battery-operated, generate sporadic traffic, and need not follow any periodic schedule. Accordingly, on-demand request originated transmissions can allow these devices to stay in sleep mode for a very long period and only wake-up when requested for a response. It can help in reducing the energy consumption of an IoT device.
\subsubsection{Clustering}
Clustering is a mechanism that allows devices to communicate with a BS through a cluster head (CH). A CH collects data from cluster members (CM) and forwards the aggregated data to the BS. The CH can be a fixed or a randomly chosen device, to distribute the energy load on all the devices \cite{lung2010using}. This process restricts CMs to forward their data only to the CH, thus saving the transmission power and channel resources. Clustering can be done for various purposes, including scalability, load balancing, reducing routing delay, collision avoidance, and EE routing. Clustering algorithms are generally designed based on probabilistic (random or hybrid) and deterministic (fuzzy-based, weight-based, or compound-based) approaches \cite{afsar2014clustering}. Multiple algorithms have been proposed for clustering, including hybrid, EE, and distributed clustering \cite{younis2004heed}, power-efficient and adaptive clustering \cite{yi2007peach}, and link-aware clustering mechanism \cite{wang2013lcm}. These algorithms provide 25\%-50\% reduction in energy consumption of an IoT device.
\subsubsection{Power Saving Mode (PSM)} 
In this mode, a device remains connected to the network even during the sleep mode. Therefore, the energy required to establish a new connection for data transmission can be saved. When a device is in the PSM, it can only be reached for device termination services after a periodic tracking area update. Reduction in energy consumption using the PSM comes with a cost of transmission latency, which makes it less attractive for most of the c-IoT applications. There are some advancements in the PSM mode that provide desired delay performance keeping the energy consumption to a minimum, such as a smart PSM (SPSM) \cite{qiao2005smart}.
\subsubsection{Energy-Efficient Scheduling and Scheduled Rendezvous}
An efficient network scheduling mechanism that can adapt to the network conditions is necessary for massive IoT networks. It can not only increase the number of connections in the network but also reduce the energy consumption of the connected devices. Network scheduling is done in either a centralized or a distributed way. The authors in \cite{wang2006survey} performs a survey on the EE scheduling mechanism for sensor networks. The authors have concluded that choosing an efficient scheduling mechanism depends on many factors, including connected devices' capabilities, network structure, and deployment strategy. One of the examples of scheduling an IoT network is scheduled rendezvous \cite{al2010energy}. In this type, devices adjust their wake/sleep schedules according to the rendezvous time. The advantage of this type of MAC protocol is that the network knows when a device wakes up, and it assigns the network resources accordingly. This type of scheduling can reduce the energy consumption of a device because it schedules the device's wake up only when the network expects to receive or send data to that device.  

The above-discussed techniques have been used in various IoT applications for energy conservation. However, most of these techniques cause communication delays and poor QoS \cite{sheng2018toward}. There is always a trade-off between energy conservation and QoS; thus, necessary optimization is required to achieve a specific QoS target for any IoT application. Improving energy efficiency and providing sufficient QoS are the basic requirements for designing a MAC  protocol for future IoT applications. Moreover, with the introduction of new computing paradigms (cloud and fog), IoT devices can spend most of their life in sleep/idle mode by employing duty cycling, and on-demand request originated transmissions. We have also identified that in low traffic load conditions, the power consumed during sleep mode becomes a determining factor for the IoT device's battery lifetime. Furthermore, we raise the following questions that might help researchers to find potential research problems.
\begin{itemize}
 \item Is adaptive clustering more energy-efficient when compared to static clustering? What kind of adaptive clustering is more suitable for a long battery lifetime (location-based, content-based, grid-based, or device' type-based)?
 \item Can prior estimation of traffic and channel fluctuations reduce the power consumption of an IoT device? If yes, then where (device or AP) and how to do it?
 \end{itemize}

Next, we will discuss MAC designs for cellular IoT connectivity in terms of their energy consumption.

\begin{table*}
\centering
\caption{Different Features of MTC \& H2H Communication}
\label{tb_3}
\resizebox{\textwidth}{!}{\begin{tabular}{|l|l|l|}
\hline
\textbf{Features}     & \textbf{Machine-Type Communication (MTC)}     & \textbf{Human-to-Human Communication (H2H)}                                                              \\ \hline
\textbf{Message Size}       & \begin{tabular}[c]{@{}l@{}}Small amount of data (for sensing purpose)\end{tabular}      & \begin{tabular}[c]{@{}l@{}}Large amount of data (Multimedia streaming, etc.)\end{tabular}              \\ \hline
\textbf{Traffic Generation}                                                       & Huge UL data, Small DL data                                                                                    & Small UL data, Huge DL data                                                                              \\ \hline
\textbf{\begin{tabular}[c]{@{}l@{}}Transmission Periodicity\end{tabular}}       & \begin{tabular}[c]{@{}l@{}}Rapidly change according to the requirements (Sporadic traffic)\end{tabular}      & \begin{tabular}[c]{@{}l@{}}Asynchronous \& random in nature\end{tabular}                               \\ \hline
\textbf{\begin{tabular}[c]{@{}l@{}}Connection \& Access Delay\end{tabular}}    & \begin{tabular}[c]{@{}l@{}}Moderate (Application-oriented)\end{tabular}                                      & \begin{tabular}[c]{@{}l@{}}Demanding (Connection delays are not tolerated)\end{tabular}                \\ \hline
\textbf{Mobility}                                                                 & \begin{tabular}[c]{@{}l@{}}Generally devices are fixed (Not a big concern)\end{tabular}                      & Always required                           \\ \hline
\textbf{Control Signaling}                                                        & \begin{tabular}[c]{@{}l@{}}Very small (Signaling required for connection establishment only)\end{tabular} & \begin{tabular}[c]{@{}l@{}}Very high (For reliable and long-time connection)\end{tabular}               \\ \hline
\textbf{\begin{tabular}[c]{@{}l@{}}Lifetime \& Energy Efficiency\end{tabular}} & \begin{tabular}[c]{@{}l@{}}Very critical (Operate in remote areas for years)\end{tabular}                    & \begin{tabular}[c]{@{}l@{}}Energy may not be a critical issue (Humans can charge devices)\end{tabular} \\ \hline
\textbf{\begin{tabular}[c]{@{}l@{}}No. of devices connected\end{tabular}}       & \begin{tabular}[c]{@{}l@{}}Very large (Ericsson predicted 50 billion \cite{ericsson2011more})\end{tabular}           & \begin{tabular}[c]{@{}l@{}}Large (The Independent predicted 7 billion )\end{tabular}   \\ \hline
\textbf{Device Duty Cycle}                                                        & \begin{tabular}[c]{@{}l@{}}Generally 1\% (Mostly in sleep mode)\end{tabular}                                & 100\%                                                                                                   \\ \hline
\end{tabular}}
\end{table*}
\section{Energy-Efficient MAC Design for Cellular IoT Networks}
As the need for a global IoT network increasing rapidly, the unlicensed band operating techniques cannot fulfill the requirements of a diverse IoT network alone. To address this issue, 3GPP has standardized MTC communication in Release 11 of the LTE standard. The advantage of using cellular networks is that it can provide ubiquitous coverage, mobility, roaming, security, and lower deployment cost. It makes cellular IoT a favorable infrastructure-based solution for global IoT networks. Moreover, the cellular community has a mature ecosystem that enables industry partnerships for future developments. Cellular networks were primarily designed for H2H communication, and integration of MTC might raise new challenges. QoS and traffic requirements for H2H communication and MTC are different from each other. H2H communication requires high throughput, delay sensitivity, and efficient mobility management, whereas MTC is generally delay-tolerant and needs low data rate for sporadic data transmission. Table \ref{tb_3} provides a detailed comparison of MTC and H2H communication types.

\begin{figure}[ht]
\centering
{\includegraphics[width=3.3in]{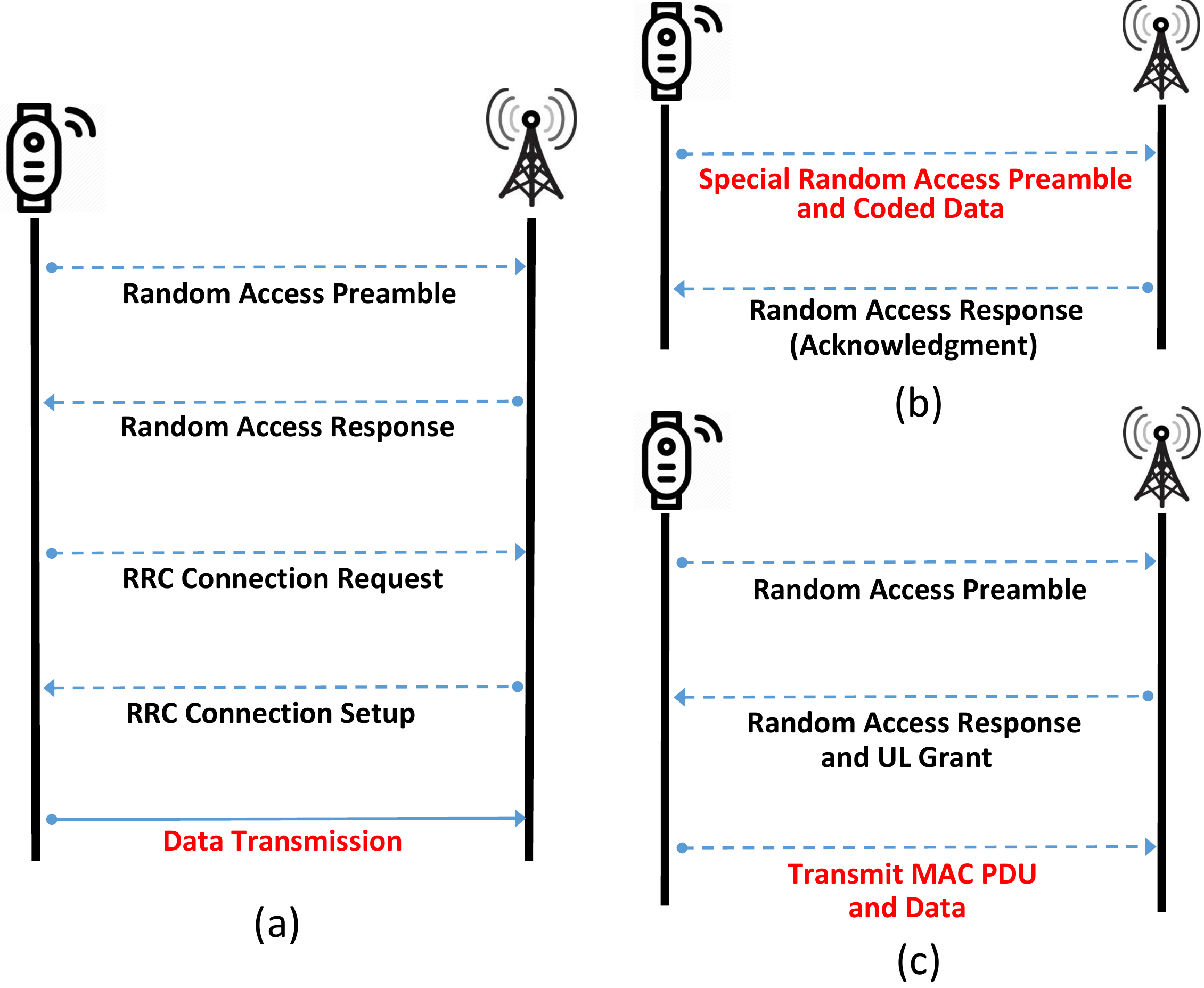}}
\caption{(a) Physical random access channel of LTE (b) Sending data with preamble message (c) Sending data with connection request message.}
\label{randomaccess}
\end{figure}

Major energy consumers for MTC communication in conventional cellular networks are large control signaling overhead and large duty cycling \cite{azari2016energy}. In a cellular network, MTC devices contend to access the channel using a physical random access channel (PRACH), as shown in Fig. \ref{randomaccess}(a). The normal RACH process, which was initially designed for H2H communication, requires a large control signaling, making it power-hungry and inefficient for MTC traffic \cite{laya2014random}. To reduce the duty cycling and for optimizing the power consumption, the PSM and various discontinuous reception (DRx) mechanisms were introduced. Moreover, there are various other challenges associated with massive MTC in cellular networks. The readers are referred to \cite{biral2015challenges}. It is important to update the conventional cellular networks to address these challenges. The 3GPP also highlights the importance of improving the conventional cellular networks according to the MTC requirements, as current cellular networks are not feasible to accommodate IoT traffic \cite{3GPP}.

There are improvements proposed in the RACH and the DRx mechanism for MTC traffic. We will discuss these improvements in the context of their energy efficiency below.

\subsection{Optimized MAC}
A conventional RACH process contains four control messages for establishing a radio connection, as shown in Fig. \ref{randomaccess}(a). This mechanism is power-hungry for MTC traffic due to its large control signaling. To reduce the control signaling overhead, the authors of \cite{chen2010machine} propose the idea of an optimized MAC. In this protocol, data transmission is possible without establishing a radio resource control (RRC) connection. This protocol embeds data in either preamble or the connection request message, as shown in Figs. \ref{randomaccess}(b) and \ref{randomaccess}(c). Therefore, devices do not need to support the complex RRC protocol and thus, save the network resources and enhance energy efficiency.

\subsection{MAC Design Using Partial Clustering}
\begin{figure}[ht]
\centering
{\includegraphics[width=3.3in]{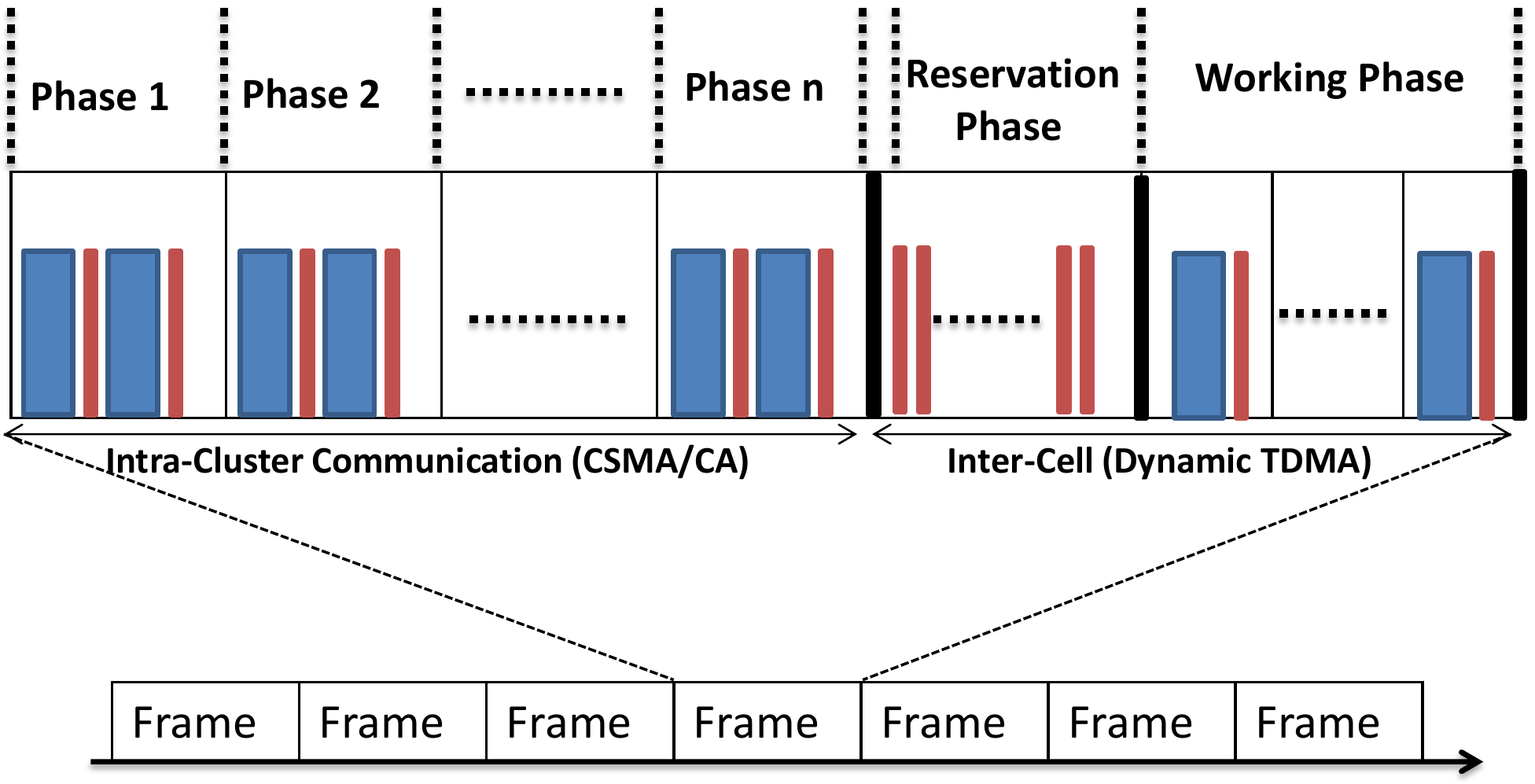}}
\caption{Frame structure for cluster-based MAC protocol \cite{azari2014energy}.}
\label{frame_structure}
\end{figure}

The authors of \cite{azari2014energy} propose the idea of partial clustering. In this mechanism, intra-cluster communication uses a contention-based protocol, and communication between CHs and the BS uses a reservation-based protocol. As clustering is a tool to decrease the number of access requests to the BS, this mechanism makes the overall system scalable. However, when the BS has the same distance to CMs and CH, energy consumption due to two-hop communication becomes critical. To address this, the authors propose to use clustering only when MTC devices are far from the BS. The frame structure for transmission is divided into two subframes, as shown in Fig. \ref{frame_structure}. The use of two subframes makes it compatible with existing cellular network architectures. The inter-cluster subframe is further divided into a reservation phase and a working phase. This idea of two-hop transmission is designed for energy efficiency in situations where multiple short-lived sessions are available.

\subsection{Discontinuous Reception (DRx) Mechanism}
\begin{figure}[!h]
\centering
{\includegraphics[width=3.3in]{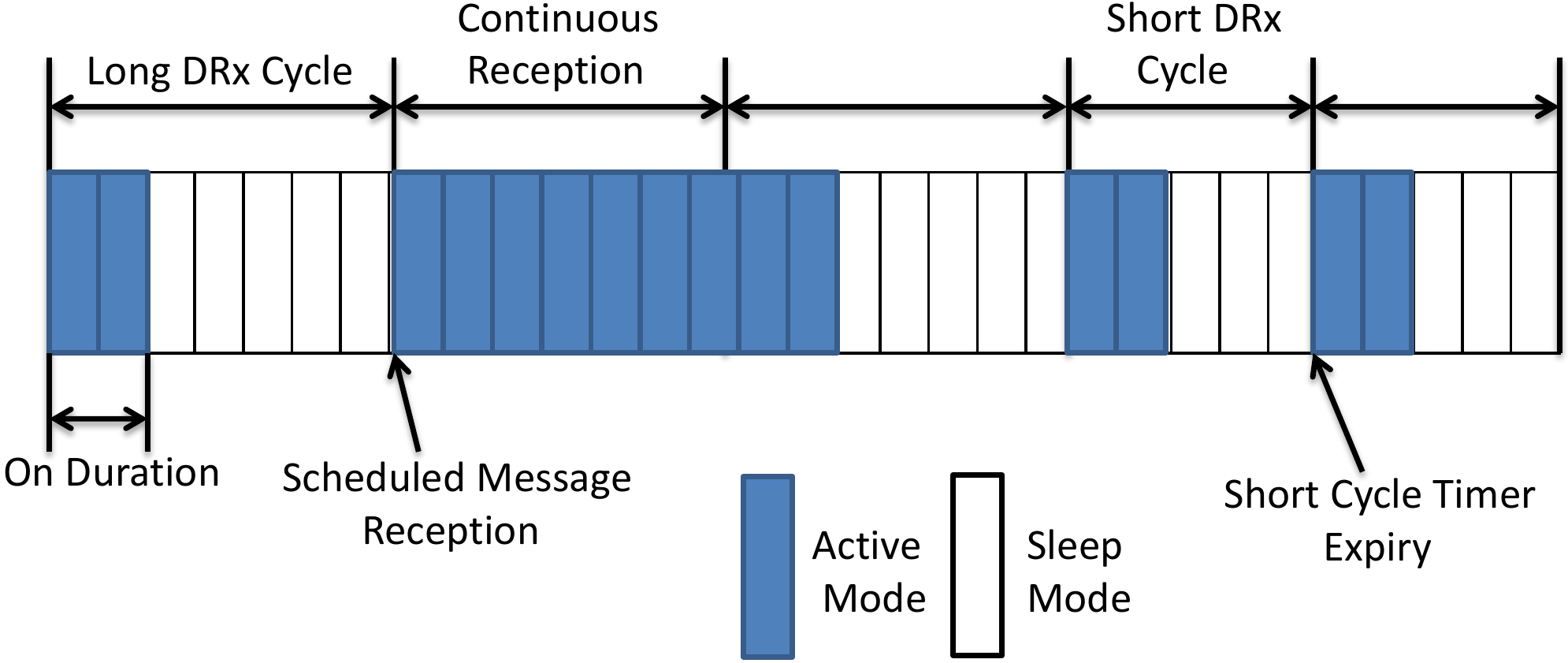}}
\caption{Discontinuous reception (DRx) mechanism in LTE \cite{zhou2013lte}. }
\label{DRx_mechanism}
\end{figure}

MTC devices generate sporadic data traffic and the time of active data transmission is very small. Some devices (e.g., temperature sensing devices) generate only a few bytes of data in a whole day; thus, keeping them active all the time is inefficient. For this purpose, the DRx mechanism was introduced for MTC communication in cellular networks \cite{bontu2009drx}. In this mechanism, MTC devices turn their radio on periodically to listen to the physical downlink control channel (PDCCH) for incoming traffic, as shown in Fig. \ref{DRx_mechanism}. There are two states in the DRx mode; RRC-idle state and RRC-connected state. In the RRC-idle, the BS releases the RRC connection of the MTC device and requests the mobility management entity (MME) to remove the MTC device s1 connection(the connection between eNodeB and extended packet core (EPC)). Although the device kept registered with the MME, it does not have an active session with MME. Therefore, the device itself is responsible for mobility management, and the network is unaware of its movements during that period. For initiating UL traffic, the MTC device needs to request its serving BS for an RRC connection. For the DL traffic, the BS either page the device or wait for the device to wake-up. In the RRC-connected state, the MTC device does not terminate its connection with the network. When the MTC device is idle, it may request the initiation of the DRx mode. When a packet arrival is detected, the MTC device terminates the DRx mode and returns to the active mode for packet reception. In this state, the network keeps track of the device, and mobility management is not the device's responsibility. By employing the DRx mechanism with a suitable configuration, the power consumption of a device can be significantly reduced without affecting the QoS requirement \cite{mihov2010analysis}.

The conventional DRx mechanism enhances the energy efficiency of an MTC device at the cost of increased packet delay. Because if the receiving device is in the DRx mode, then the packet destined for it is stored at the BS until that device listens to PDCCH for its DL traffic. There are multiple schemes \cite{karthik2013practical,yu2012traffic,tseng2016delay} proposed for DRx parameters configuration based on traffic activity in the network to ensure power efficiency along with reduced packet delay. There are also some further improvements in the DRx mechanism, which can enhance energy efficiency even further. Table \ref{tb_2424} provides a brief summary of these improvements.

\begin{table*}
\centering
\caption{Discontinuous Reception (DRx) Mechanism for Power Saving and Latency Reduction}
\label{tb_2424}
\resizebox{\textwidth}{!}{\begin{tabular}{|l|l|l|l|}
\hline
\textbf{References}     & \textbf{Model}                                                                                 & \textbf{Task}                                                                 & \textbf{Brief Summary}                                                                                                                                                                                                                                                                                                         \\
\hline
Hsieh {\emph{et al.}} \cite{hsieh2016accurate}   & Semi-Markov   & Power Saving  & \begin{tabular}[c]{@{}l@{}}Semi-Markov process is used to analytically model DRx mechanism and fixed\\sleep duration is replaced with variable sleep duration. High power saving is a-\\hieved at the expense of large buffering delay.\end{tabular}                                                                  \\
\hline
Liu {\emph{et al.}} \cite{liu2013adaptive}     & \begin{tabular}[c]{@{}l@{}}Counter Driven Adaptive DRx\\ (CDA-DRx)\end{tabular}                & Power Saving                                                                  & \begin{tabular}[c]{@{}l@{}}Reduction in signaling overhead of DRx mechanism using counters at both en-\\ds(UE and eNodeB). A considerable decrease in power consumption is achiev-\\ed keeping the average latency same.\end{tabular}                                                                                       \\
\hline
Jha {\emph{et al.}} \cite{jha2013adaptive}     & Semi-Markov                                                                                    & \begin{tabular}[c]{@{}l@{}}Power Saving +\\ Latency Reduction\end{tabular}  & \begin{tabular}[c]{@{}l@{}}Tuning of DRx parameters is done to optimize balanced power consumption\\ and average latency. Authors also stated that with the knowledge of mean\\inter-arrival time, benefits can be extracted from DRx operation in terms of\\balance between power consumption and latency.\end{tabular} \\
\hline
Sergio {\emph{et al.}} \cite{herreria2015adaptive}    & Adaptive DRx                                                                                    & Power Saving                                                                  & \begin{tabular}[c]{@{}l@{}}It allows an adjustable DRx cycle according to the network requirements.\\It allows devices to remain in the PSM for longer period of time; thus,\\ help reducing their energy consumption.\end{tabular}                                                                \\
\hline
Balasubramanya {\emph{et al.}} \cite{balasubramanya2016drx}    & DRx with Quick Sleep                                                                     & Power Saving & \begin{tabular}[c]{@{}l@{}}It uses quick sleep indication technique, in which bits are allocated to each\\group of devices that indicated the upcoming paging request from the BS. It\\provides 66\% better energy efficiency as compated to conventional DRx. \end{tabular}                              \\
\hline
Yu {\emph{et al.}} \cite{yu2012traffic}      & \begin{tabular}[c]{@{}l@{}}Partially Observable Markov\\ Decision Process (POMDP)\end{tabular} & Power Saving                                                                  & \begin{tabular}[c]{@{}l@{}}Traffic-based DRx cycle adjustment (TDCA) scheme has been proposed to\\adaptively control DRx cycles for decreasing signaling overhead. It enhances\\energy consumption efficiency keeping QoS intact.\end{tabular}                                                                          \\
\hline
Fowler {\emph{et al.}} \cite{fowler2015analytical}    & Extended DRx (eDRx)                                                                                    & Power Saving & \begin{tabular}[c]{@{}l@{}}It allows a device to extend its sleep cycle upto 60 frames each with 10.24 s\\ before it responds to a paging request from the BS. The device may or\\may not be allowed to adopt eDRx based on network configuration.\end{tabular}                                               \\
\hline
Karthik {\emph{et al.}} \cite{karthik2013practical} & Discrete Probability Model                                                                     & Power Saving                                                                  & \begin{tabular}[c]{@{}l@{}}A pragmatic algorithm is proposed for selecting suitable DRx configuration ino\\rder to gain high power efficiency with tolerable delay constraints.\end{tabular}                                                                                                                              \\
\hline
Wang {\emph{et al.}} \cite{wang2015optimizing}    & \begin{tabular}[c]{@{}l@{}}Truncated Pareto Distributed\\ Model\end{tabular}                   & \begin{tabular}[c]{@{}l@{}}Power Saving +\\ Latency Reduction\end{tabular} & \begin{tabular}[c]{@{}l@{}}Online power saving scheme (OPSS) is proposed for DRx mechanism using\\estimation \& optimization phases. Effect of various traffic conditions is presented.\\ It outperforms legacy DRx schemes both in packet delay and power consumption.\end{tabular}
\\
\hline
\end{tabular}}
\end{table*}

Next, we discuss the EE cellular standards designed specifically for IoT applications.
\section{Energy-Efficient Cellular Standards for IoT}
IoT applications are diverse, and almost every other application requires a distinct technological demand. Therefore, the ``one solution fits all'' approach does not necessarily fulfill the requirements of diverse IoT applications.  To help address this issue, 3GPP proposed three standards, especially for IoT applications, in its Release 13, which are extended-coverage GSM (EC-GSM), narrow-band IoT (NB-IoT), and long term evolution for machines (LTE-M). Their design is based on the power consumption, QoS, and latency requirements of IoT applications. We discuss these standards in terms of their power consumption in this section. We also provide a technical comparison of these standards with other proprietary LPWA technologies in Table \ref{lpwa_cell}.

\subsection{LTE-M}
It is a low-cost and low-power standard proposed by 3GPP in its Release 12 for IoT applications \cite{14fghfch}. LTE-M is a solution for IoT applications that require low-power consumption, reliability, low-latency, and a large coverage area. It can be deployed in the conventional LTE system with only a software upgrade, which makes it a very promising solution in terms of fast and low-cost deployments \cite{dawaliby2016depth}. LTE-M utilizes lesser bandwidth (1.4 MHz) as compared to the conventional LTE (20 MHz). Data rates for its UL and DL channels are also reduced to less than 1 Mbps, which makes it suitable for m-IoT applications only. It uses two different MAC layer protocols for channel access. For m-IoT application, it utilizes contention-based channel access in which frame-slotted ALOHA (FSA) is used. FSA is used for initial channel access because thousands of devices try to access the channel, and allocating separate resources is not feasible. On the other hand, for c-IoT applications, contention-free channel access is used, to provide mobility management and seamless handover. For contention-free channel access, frequency division multiple access (FDMA) and time slot reservation mechanism is used \cite{laya2014random}. To ensure less power consumption, LTE-M employs different mechanisms such as PSM, eDRx, and C-DRx. These mechanisms allow IoT devices to go to sleep mode by turning their radio off while keeping their connection intact with the core network for allowing paging \cite{ratasuk2015overview}. The current consumption in a sleep mode can be as low as 4 $\mu A$ and reaches up to 320 $mA$ in an active mode.

\subsection{EC-GSM}
EC-GSM, also known as EC-eGPRS, is an attempt to utilize legacy second-generation (2G) cellular networks for low-power, low data rate connectivity. 2G cellular networks (GPRS/ eGPRS), which operate in multiple frequency bands, such as 850, 900, 1800, and 1900 MHz. EC-GSM based on 2G cellular networks can become a global infrastructure-based solution for IoT connectivity, as they are already deployed in most parts of the world. This advantage over other cellular networks-based solutions can make it a feasible, easy, and cheaper solution for global IoT integration. Moreover, with the introduction of added features of the EE operation, and improved coverage by almost 20 dB over eGPRS, it is also suitable for the energy-constrained IoT applications. These newly added features can be integrated into existing GSM networks using only software upgrades to the core and radio access networks. 3GPP has also issued new technical features for EC-GSM in its Release 13. After the integration of these features, EC-GSM can achieve a maximum of 240 kbps data rate on its UL channel with a total bandwidth of 600 kHz, compared to GSM bandwidth of 2.4 MHz. Each channel of the EC-GSM is allocated a bandwidth of 200 kHz, and both the UL and the DL channels support TDMA and FDMA-based transmissions. EC-GSM uses PSM and eDRX mechanisms for the EE operation, which makes it less power-hungry compared to the conventional GSM. Some changes at the MAC layer are required to implement these new features, such as the introduction of new MAC layer messages (for the extended coverage-packet data channel), changes in the RAN (to support paging in eDRX), and reduction in stringent requirement of neighbor cell monitoring (for power-efficient operation).

\begin{table*}[ht]
\centering
\caption{Proprietary Solutions vs. Cellular Standards for IoT}
\label{lpwa_cell}
\resizebox{\textwidth}{!}{\begin{tabular}{|c|c|c|c|c|c|c|}
\hline
\textbf{Technology}                                                        & \textbf{LoRaWAN}\cite{ie32e22iee}                                                    & \textbf{SigFox} \cite{ieee22iee}                                                                        & \textbf{RPMA} \cite{ie589922iee}                                                                      & \textbf{NB-IoT}                                                                                                            & \textbf{EC-GSM}                                                  & \textbf{LTE-M}                                                                  \\ \hline
\textbf{Spectrum}                                                          & \begin{tabular}[c]{@{}c@{}}Un-licensed\\ \textless1 GHz\end{tabular} & \begin{tabular}[c]{@{}c@{}}Un-licensed\\ \textless1GHz\end{tabular}                    & \begin{tabular}[c]{@{}c@{}}Un-licensed\\ 2.4 GHz\end{tabular}                       & \begin{tabular}[c]{@{}c@{}}Licensed\\(1) Stand alone in GSM band\\(2) LTE Guard band\\(3) In-band LTE\end{tabular} & \begin{tabular}[c]{@{}c@{}}Licensed\\ In GSM band\end{tabular}   & \begin{tabular}[c]{@{}c@{}}Licensed\\In-band LTE\end{tabular}                  \\ \hline
\textbf{\begin{tabular}[c]{@{}c@{}}Peak data rate \\ (UL/DL)\end{tabular}} & 50 kbps/290 bps                                                       & 100 bps/600 bps                                                                          & 624Kbps/156Kbps                                                                    & \begin{tabular}[c]{@{}c@{}}204.8 kbps (Multi-tone)\\ 20 kbps (single-tone)\\ /234.7 kbps\end{tabular}                     & 240 kbps/240 kbps                                            & \textless1 Mbps/\textless1 Mbps                                                   \\ \hline
\textbf{Peak current}                                                      & 32 $m$A                                                                & 11 $m$A                                                                                   & 30 $m$A                                                                               & 120 $m$A -- 300 $m$A                                                                                                                &                          120 $m$A -- 320 $m$A                                        & 320 $m$A                                                                           \\ \hline
\textbf{Sleep current}                                                     & $1 \mu$A                                                                 & $1\mu$A                                                                                    & $1 \mu$A                                                                                & $5 \mu$A                                                                                                                        & $4 \mu$A                                                              & $4 \mu $A                                                                             \\ \hline
\textbf{Transmit power}                                                    & 2 dBm-20 dBm                                                          & \begin{tabular}[c]{@{}c@{}}14dBm (Compliant\\with local constraints)\end{tabular} & 22 dBm                                                                              & 23 dBm                                                                                                                      & 23 dBm -- 33 dBm                                                      & \begin{tabular}[c]{@{}c@{}}23 dBm:Class 3 UE\\ 20 dBm:Class 5 UE\end{tabular}     \\ \hline
\textbf{Power saving}                                                      & LoRaWAN                                                             &    \begin{tabular}[c]{@{}c@{}}Deep sleep\\ mechanism\end{tabular}                                                                                   & \begin{tabular}[c]{@{}c@{}}Deep sleep\\ mechanism\end{tabular}                                                                                    & PSM + eDRx                                                                                                                 & PSM + eDRx                                                       & \begin{tabular}[c]{@{}c@{}}PSM + eDRx\\ C-DRx\end{tabular}                      \\ \hline
\textbf{Power efficiency}                                                  & High                                                                & Very High                                                                              & High                                                                               & Moderate                                                                                                                   & Moderate                                                         & Moderate                                                                        \\ \hline
\textbf{MAC protocols}                                                     & \begin{tabular}[c]{@{}c@{}}Un-slotted\\ ALOHA\end{tabular}          & \begin{tabular}[c]{@{}c@{}}Un-slotted\\ ALOHA\end{tabular}                             & CDMA-like                                                                          & \begin{tabular}[c]{@{}c@{}}SC-FDMA\\ CDMA\end{tabular}                                                                     & \begin{tabular}[c]{@{}c@{}}TDMA\\ FDMA\end{tabular}              & \begin{tabular}[c]{@{}c@{}}SC-FDMA\\ TDMA, FSA\end{tabular}                    \\ \hline
\textbf{Range (km)}                                                        & \begin{tabular}[c]{@{}c@{}}Rural:15\\ Urban:2-5\end{tabular}        & \begin{tabular}[c]{@{}c@{}}Rural:30-50\\ Urban:3-10\end{tabular}                       & \begin{tabular}[c]{@{}c@{}}Rural:30-50\\ Urban:1-5\end{tabular}                    & \begin{tabular}[c]{@{}c@{}}Rural:10-15\\ Urban:1-2\end{tabular}                                                            & \textless25                                                      & 7-10                                                                            \\ \hline
\textbf{Deployment status}                                                 & \begin{tabular}[c]{@{}c@{}}Commercially\\ deployed\end{tabular}     & \begin{tabular}[c]{@{}c@{}}Early deployment \\ in the US \& the EU\end{tabular}                & \begin{tabular}[c]{@{}c@{}}Commercially\\ deployed\end{tabular}                    & \begin{tabular}[c]{@{}c@{}}Commercially\\ deployed\end{tabular}                                             & \begin{tabular}[c]{@{}c@{}}Commercially \\ deployed\end{tabular} & \begin{tabular}[c]{@{}c@{}}Commercially \\ deployed\end{tabular}                                                                  \\ \hline
\textbf{Latency}                                                           & \begin{tabular}[c]{@{}c@{}}Insensitive to \\ latency\end{tabular}   &  upto 6 s                                                                                      &                 2 s -- 4 s                                                                   & 1.6 s -- 10 s                                                                                                                   & 700 ms -- 2 s                                                         & 10 ms -- 15 ms                                                                       \\ \hline
\textbf{\begin{tabular}[c]{@{}c@{}}Battery \\ lifetime\end{tabular}}           & 10 years                                                            & \begin{tabular}[c]{@{}c@{}}20 years (2.5 Ah)\\ 3 messages per day\end{tabular}       & \begin{tabular}[c]{@{}c@{}}21.8 years (2.5 Ah)\\ 200 kB data per day\end{tabular} & \begin{tabular}[c]{@{}c@{}}19 years (2.5 Ah)\\ 1 message per day\end{tabular}                                            & 10 years (5Ah)                                                   & \begin{tabular}[c]{@{}c@{}}19 years (2.5 Ah)\\ 1 message per day\end{tabular} \\ \hline
\textbf{\begin{tabular}[c]{@{}c@{}}Link Budget/\\ Coverage\end{tabular}}   & 157 dB                                                               & 149 dB                                                                                  & 177 dB                                                                              & 164 dB                                                                                                                      & 164 dB                                                            & 155 dB                                                                           \\ \hline
\end{tabular}}
\end{table*}

\subsection{NB-IoT}
NB-IoT is another 3GPP standard for IoT, which ensures minimized signaling overhead and improved battery life \cite{wang2017primer}. It was released by 3GPP in its Release 13. It can be deployed in three different scenarios;  stand-alone in GSM band with 200 kHz channel, in-band LTE with 180 kHz channel, and guard-band of LTE \cite{rico2016overview}. In early versions of NB-IoT, it uses the same channel access mechanism as LTE to establish an RRC connection \cite{mwakwata2019narrowband}. However, in later versions, a new single-tone frequency hopping random access (NPRACH) procedure has been introduced \cite{lin2016random}. In NPRACH, both single-tone and multi-tone frame structures can be used for UL transmission. In a single-tone frame structure, only one subcarrier with a frequency spacing of 15 kHz and 3.75 kHz is available that allows 12 and 48 different preambles, respectively. On the other hand, in a multi-tone frame structure, the number of subcarriers can be 3, 6, or 12 with 15 kHz frequency spacing \cite{ingfghfch}. The single-tone UL data rate can reach up to 20 kbps, whereas the multi-tone UL data rate can be as high as 204.8 kbps \cite{ratasuk2016nb}.  For reduced power consumption, NB-IoT utilizes PSM and eDRx mechanism. The eDRX mechanism is supported in RRC-idle and RRC-connected modes. To keep a device registered with the network, it supports a paging cycle of 10.24 s and 3 h in RRC-connected and RRC-idle modes, respectively \cite{bontu2009drx}. It supports the highest eDRx cycle length as compared to EC-GSM and LTE-M. The authors in \cite{lee2017prediction} propose a prediction based EE channel access mechanism for NB-IoT. In this mechanism, prediction-based resource allocation is done. This mechanism predicts the processing delay and UL occurrences by examining the transmitted packet. Moreover, the BS pre-assigns the channel resources for the UL transmission without a scheduling request process. The authors claimed a 34\% reduction in energy consumption when compared to the conventional NB-IoT channel access method. There are some limitations of NB-IoT as well. It does not support QoS, as its traffic is the best effort, and only half of its messages are acknowledged. It also does not ensure ultra-reliability and low latency (it can reach up to 10 s), which makes it less suitable for c-IoT applications, such as traffic safety and control, smart grid, and other industrial applications \cite{raza2017low}.

\section{Energy-Efficient Massive MTC (mMTC) in 5G and Beyond}
In conventional cellular networks, the focus of the network providers is to provide reliable and high data rate connectivity. The need to ensure this type of communication has led to sophisticated and complex MAC layer designs that require large control signaling. Control signaling overhead is only admissible when it is small compared to the payload data \cite{shah2019protocol}. Because most of the IoT applications generate a small amount of data, conventional MAC layer protocols are inefficient for MTC in cellular networks. This phenomenon creates new challenges for designing MAC layer protocols for 5G and beyond cellular network. 5G introduces mMTC to accommodate a large number of small data-generating devices \cite{chen2020massive}. There are some underline requirements for 5G and beyond that need to be addressed while designing MAC layer protocols. Some of these requirements are: a BS should support a large number of devices (up to 300,000); small packet size; large UL traffic as compared to DL traffic; support sporadic traffic generation and EE transmission \cite{agiwal2018towards}.

Several research groups are actively working on the EE MAC designs for mMTC in 5G and beyond cellular networks \cite{miuccio2020joint,hou2019interference,8878115,mohammadkarimi2018signature,liu2020energy}. One of the solutions is the one-shot random access, in which a device sends data right after it wakes up without any synchronization and coordination procedure \cite{wunder20145gnow}. Although this procedure is EE, it poses multiple challenges, including a high collision rate due to its asynchronous nature, which limits the channel throughput. This issue of limited throughput is addressed in \cite{wunder2015compressive} using compressive coded random access mechanism. The authors propose that using a common overloaded control channel makes the one-shot random access procedure possible. The authors in \cite{bockelmann2016massive} use the UL sparse code multiple access enabled contention-based mechanisms to overcome the problems of scalability and network overload. Their simulation results show that their proposed solution can accommodate three times more MTC devices as compared to the contention-based orthogonal frequency division multiple access. The same authors presented another solution using coded random access (CRA) coupled with compressed sensing-based multiuser detection. The idea of CRA is similar to the frameless ALOHA, in which devices independently contend for the channel on the slot basis, with a fixed prior channel access probability, without any coordination between devices. This solution can significantly enhance scalability, energy efficiency, and channel resource allocation.

The evolution of 5G New Radio (NR) also paves the way for the EE Industrial IoT and URLLC. The 3GPP has finalized its release 16 and release 17, which includes a mechanism for the wake-up signal for the IoT devices \cite{peisa20205g}. By using this specific wake-up signal, an IoT device can further reduce its power consumption \cite{3gpp2020}. It also offers enhancements to the legacy scheduling and control signaling mechanisms of the cellular networks to efficiently accommodate mMTC communication. Some other enabling technologies of 5G and beyond cellular networks can also help establish EE mMTC communication. Device-to-Device (D2D) communication is one of those technologies. In D2D communication, devices present nearby communicate directly without routing their data through the BS \cite{shah2019impact}. D2D communication is a suitable candidate for reducing the energy consumption of IoT devices due to its reduced transmission power levels. It also enables reuse of the cellular spectrum; thus helps to reduce the spectrum scarcity issue. The authors of \cite{liang2018energy} propose an EE DRX scheduling for D2D users. Their scheme optimizes the sleep operation of the D2D users using the DRx mechanism while maintaining the required QoS guarantees. This scheme also maintains the transmission and interference relationship graph to mitigate the effects of D2D interference, which leads to better channel quality. Moreover, 5G and beyond cellular networks envision to allow licensed and unlicensed interoperable communication. IEEE 1932.1 working group has proposed a standard for licensed and unlicensed spectrum interoperability. This standard includes an interoperation mechanism among MAC protocols designed for licensed and unlicensed spectrum operations and a controller for devices' coordination. The authors of \cite{shah2020statistical} propose licensed and unlicensed interoperable D2D communication. The proposed scheme not only enhance the spectral efficiency of the D2D devices, but it also enables massive connectivity using an unlicensed network while keeping the advantages of the cellular networks. To this end, the authors of \cite{ismaiel2017scalable} propose a scalable MAC protocol that offloads the D2D users from the cellular network to the unlicensed network. This protocol uses the point coordination function access mechanism to manage D2D users in a dense environment. 

Similar to D2D communication, Unmanned aerial vehicle (UAV)-assisted MTC is also gaining popularity in 5G-NR. In this type of communication, UAVs/drones are used for data collection, which is one of the most critical tasks of MTC. UAVs (data collectors) can complement cellular BSs and process MTC traffic to reduce network congestion. Although UAV-assisted MTC communication brings many benefits \cite{li2020uav}, there are challenges associated with spectrum sharing in cellular networks. The authors in \cite{9103583} studies the spectrum sharing between MTC devices and cellular users in the presence of UAVs, which also use the cellular spectrum for communication. The authors propose a time-division duplexing protocol and provide a comparison with state-of-the-art orthogonal and nonorthogonal spectrum sharing protocols. Further, the paper also investigates the optimized transmit powers for EE operation of devices connected in the network. Similarly, the authors in \cite{wang2019energy} aim to reduce the energy consumption of MTC devices by jointly optimizing devices' transmission schedule over time and the UAV trajectory. In short, UAV-assisted MTC can play an important role in m-MTC in future air-ground integrated IoT networks. To conclude, EE and massive connectivity is still an open research challenge in 5G and beyond cellular networks. The readers are referred to \cite{chen2020massive} and the references therein for further details. 
\section{Open Research Challenges}
In this section, we present some of the key research challenges for the EE MAC protocols for cellular IoT.
\subsection{Energy-Efficient Ambient Computing}
Nowadays, IoT is no longer a collection of simple sensors, rather an integration of intelligent devices. These devices can sense and gather information from their environment and respond to our needs intelligently. This yields the emergence of ambient computing. This type of computing is becoming popular recently in various domains of the IoT landscape. Although billions of sensors are deployed in various IoT applications, a small number is connected to an access network, and an even smaller number have Internet access. It is expected that billions more will be deployed soon, to provide Internet connectivity, and to manage the huge amount of data generated by these devices is a crucial task. An intelligent device must have storage and processing capabilities, high QoS, and it also needs to perform online data analytics for local decision making. To perform these tasks on an energy-constrained device is challenging and requires new research efforts to ensure reduced power consumption while supporting ambient computing. One solution is ambient backscatter communication in which an IoT device utilizes available radio frequency signals to transmit data without a battery \cite{liu2019next}. Although it provides benefits in terms of energy consumption, it has many limitations, including short-range and low data rate transmission. 
\subsection{Energy-Efficient Mobility Management}
Cellular network densification is one of the most promising solutions to overcome the capacity issue in future networks. The macro-cell based solutions are falling apart in providing access to a large number of IoT devices. This problem will even get worse with the exponential growth of connected IoT devices and their demand for high data rates and enhanced QoS. Ultra-dense multi-tier cellular networks (UDMTCNs) can provide an efficient solution to address the issue of high data rate connectivity \cite{shah2019system}. Although UDMTCNs bring in many advantages like increased network resources, enhanced throughput, decreased network load associated with a cell, and reduced transmission power of a device, it also raises several new challenges. One of the major challenges is mobility management, as the number of handovers will increase overwhelmingly in UDMTCNs. These frequent handovers can increase the signaling overhead and power consumption of a device. Several IoT applications require mobility, such as asset tracking, vehicle-to-vehicle communication, smart cities, and mobile healthcare facilities. Devices connected in these applications are battery-operated and undergo frequent handovers, which require extra power consumption. A proactive approach is required to overcome this power consumption issue caused due to high mobility. It can open doors for new research on EE mobility management solutions in future cellular networks.
\subsection{Handling Bursty MTC Traffic in Green Cellular IoT Networks}
Unlike conventional cellular network traffic, MTC traffic is bursty in nature, and therefore different mechanisms are required to accommodate such traffic. One of them is access class barring (ACB), in which traffic is categorized into different classes based on their access priorities. Several ACB mechanisms are available in the literature \cite{shah2018congestion,tavana2015congestion}, which adjusts ACB factors based on different network parameters. Nonetheless, congestion is still one of the major issues for bursty MTC traffic in cellular IoT networks. It is also important to note that traffic fluctuation and congestion at the network level also lead to more power consumption of IoT devices. To ensure a long battery life of an IoT device, new EE congestion control mechanisms for cellular IoT networks are required.
\subsection{Network-Level Energy Efficiency}
With the emergence of IoT, conventional cellular networks are lagging in providing access to the massive number of IoT devices. The issue of massive access leads to the deployment of small cells. Although a large number of small cells provides resource and EE access to a huge number of devices, it also increases the network level energy consumption. Channel access type for devices connected in a small cell plays an important role in defining the energy consumption of the cell. The channel access types can be characterized as closed, open, and hybrid. In closed access, only the authorized users can access the network, whereas open access allows any user to access the network without an authorization requirement. In the hybrid access mode, both techniques can be combined. Although it looks like open access techniques will consume more energy, there is no research available that compares the energy consumption of these access techniques and the trade-off they offer.
\subsection{Energy Harvesting}
IoT devices can be deployed in areas where replacing the batteries is economically inefficient. Many research groups are working on energy harvesting for these unattended devices to increase their operational lifetime. In \cite{coarasa2013impact}, the authors propose an idea in which mobile energy sources radiates electromagnetic energy in a pre-decided pattern, sufficient for charging an IoT device. Similarly, other research groups have also shown the advantages of energy harvesting for IoT devices \cite{ng2012energy,lin2015deepsleep}. It is an open research area with a huge potential since efficient wireless energy transfer and practical implementation of energy harvesting for IoT devices are both important open questions.

\section{Conclusions}
The influx of billions of IoT devices in recent times is a huge challenge for wireless networks. It leads to finding new wireless network solutions for providing access to these devices. Although new solutions can provide cost and resource-efficient access, energy efficiency is still a major challenge to be resolved. With the emergence of intelligent IoT devices, which need to perform storage and processing locally for data analytics, the battery lifetime of a device will be a bottleneck. To this end, we provided a comprehensive survey of the EE MAC techniques for green cellular IoT. We started with the motivation for energy efficiency in IoT networks by briefly explaining \textit{why power consumption at the MAC layer is one of the major determinants for the battery life of an IoT device.} To address this question, various energy dissipation sources at the MAC layer and their solutions were presented. Some questions were also raised in the context of low power consumption of an IoT device that can help researchers find potential research problems. Following this, various EE MAC designs for cellular IoT were presented. These designs can provide EE IoT connectivity in different use cases and applications. One of the main focuses of these designs is on the enhancement of the battery life of the IoT devices. Then, we provided a detailed discussion on the EE of the cellular standards for IoT, followed by a brief discussion on the EE massive MTC in 5G and beyond cellular networks. By looking at the recent advances in industry and possible new demanding use cases for IoT, we encourage further research on the EE massive access in 5G and beyond cellular standards for IoT that not only provides reliability but also QoS and EE connectivity.

%\section{Acknowledgments}

\appendices

\footnotesize{
\bibliographystyle{IEEEtran}
\bibliography{references}
}

\vfill\break

% Biographies are only required when submitting the final accepted version
%\begin{IEEEbiography}[{\includegraphics[width=1in,height=1.25in,clip,keepaspectratio]{waqas}}]
%{Syed Waqas Haider Shah}
%He has done his MS in Electrical Engineering from National University of Science \& Technology,Pakistan in 2016. Currently he is pursuing his PhD in Electrical Engineering form Information Technology University, Pakistan. His area of research includes Internet of Things, Machine to Machine Communication in LTE, MIMO Techniques in cellular networks and MAC protocols for MTC communication.
%\end{IEEEbiography}

\end{document}